\documentclass[letter,12pt]{article}
\usepackage{latexsym}
\usepackage{amsmath,amssymb,bm,ascmac,color,calc}
\usepackage{slashed}
\usepackage[mathscr]{eucal}
\usepackage{graphicx}
\usepackage{cite}
\usepackage{lineno}
\usepackage{jheppub}
\usepackage{multirow}
\usepackage{subfigure}
\usepackage[normalem]{ulem}
\usepackage[utf8]{inputenc}
\usepackage[T1]{fontenc}

\usepackage{color}

\usepackage{makecell}
\pdfoutput=1 

\usepackage{amsmath}
\usepackage{graphicx}
\usepackage{subfigure}
\usepackage{amssymb}
\usepackage{xcolor}
\usepackage{multirow}
\usepackage{cancel}
\usepackage{color}
\usepackage{ulem}
\usepackage{listings}
\usepackage{xcolor}
\usepackage{float}
\usepackage{slashed}

\usepackage{tikz}
\usetikzlibrary{arrows,shapes}
\usetikzlibrary{trees}
\usetikzlibrary{matrix,arrows} 				% For commutative diagram
\usetikzlibrary{positioning}				% For "above of=" commands
\usetikzlibrary{calc,through}				% For coordinates
\usetikzlibrary{decorations.pathreplacing}  % For curly braces
\usepackage{pgffor}							% For repeating patterns

\usetikzlibrary{decorations.pathmorphing}	% For Feynman Diagrams
\usetikzlibrary{decorations.markings}
\tikzset{
    vector/.style={decorate, decoration={snake}, draw},
	provector/.style={decorate, decoration={snake,amplitude=2.5pt}, draw},
	antivector/.style={decorate, decoration={snake,amplitude=-2.5pt}, draw},
    fermion/.style={draw=black, postaction={decorate},
        decoration={markings,mark=at position .55 with {\arrow[draw=black]{>}}}},
    fermionbar/.style={draw=black, postaction={decorate},
        decoration={markings,mark=at position .55 with {\arrow[draw=black]{<}}}},
    fermionnoarrow/.style={draw=black},
    gluon/.style={decorate, draw=black,
        decoration={coil,amplitude=4pt, segment length=5pt}},
    scalar/.style={dashed,draw=black, postaction={decorate},
        decoration={markings,mark=at position .55 with {\arrow[draw=black]{>}}}},
    scalarbar/.style={dashed,draw=black, postaction={decorate},
        decoration={markings,mark=at position .55 with {\arrow[draw=black]{<}}}},
    scalarnoarrow/.style={dashed,draw=black},
    electron/.style={draw=black, postaction={decorate},
        decoration={markings,mark=at position .55 with {\arrow[draw=black]{>}}}},
	bigvector/.style={decorate, decoration={snake,amplitude=4pt}, draw},
}

\tikzstyle{block} = [draw, rectangle, 
    minimum height=3em, minimum width=6em]

\newcommand{\be}{\begin{equation}}
\newcommand{\ee}{\end{equation}}
\newcommand{\beq}{\begin{equation}}
\newcommand{\eeq}{\end{equation}}
\newcommand{\bea}{\begin{eqnarray}}
\newcommand{\eea}{\end{eqnarray}}
\newcommand{\besp}{\begin{equation}\begin{split}}
\newcommand{\eesp}{\end{split}\end{equation}}

\newcommand{\Dfbd}{\mathord{\buildrel{\lower3pt\hbox{$\scriptscriptstyle\leftrightarrow$}}\over {D}_{\mu}}}

\def\0{\textbf{0}}
\def\1{\textbf{1}}
\def\2{\textbf{2}}
\def\3{\textbf{3}}
\def\4{\textbf{4}}
\def\5{\textbf{5}}
\def\6{\textbf{6}}
\def\7{\textbf{7}}
\def\8{\textbf{8}}
\def\9{\textbf{9}}

\title{Sommerfeld enhancement for puffy self-interacting dark matter}

\author[a]{Wenyu Wang,}
\author[a]{Wu-Long Xu,}
\author[b,c]{Jin Min Yang,}
\author[d]{Bin Zhu,}
\author[b,c]{Rui Zhu}

\affiliation[a]{Faculty of Science, Beijing University of Technology, Beijing 100124, P. R. China}
\affiliation[b]{CAS Key Laboratory of Theoretical Physics, 
	 Institute of Theoretical Physics, Chinese Academy of Sciences, Beijing 100190, P. R. China}
\affiliation[c]{School of Physics, University of Chinese Academy of Sciences, Beijing 100049, P. R. China}
\affiliation[d]{Department of Physics, Yantai University, Yantai 264005,  P. R. China}

\emailAdd{wywang@bjut.edu.cn}
\emailAdd{wlxu@emails.bjut.edu.cn}
\emailAdd{jmyang@itp.ac.cn}
\emailAdd{zhubin@mail.nankai.edu.cn}
\emailAdd{zhurui@itp.ac.cn}

\abstract{
We examine the Sommerfeld enhancement effect for the puffy self-interacting dark matter. We find out two new parameters to classify the self-scattering cross section into the Born, the resonance and the classical regimes for the puffy dark matter. Then we observe that the resonance peaks for the puffy dark matter self-scattering and for the Sommerfeld enhancement effect have the same locations.
Further, we find that for a large ratio between  $R_{\chi}$ (radius of a puffy dark matter particle) and  $1/m_{\phi}$ (force range), the Sommerfeld enhancement
factor approaches to 1 (no enhancement).  
Finally, for the puffy SIDM scenario to solve the small-scale
problems, the values of the Sommerfeld enhancement factor are displayed in the allowed
parameter regions. 
}

\begin{document} 
%%%%%%%%%%%%%%%%%%%%%%%%%%%%%%%%%%%
\maketitle

%\tableofcontents

%\newpage
\section{Introduction}\label{sec:intro}

In our universe over eighty percent of matter is composed of dark matter (DM) \cite{Bertone:2016nfn,Planck:2018vyg}. Beyond gravitational interaction, the nature of DM is a mystery for modern physics.  The prediction of the standard model of cosmology, i.e., the $\Lambda \rm CDM$ with cold and collisionless DM, is consistent with the observation of large-scale structures, such as the clusters of galaxies \cite{Trujillo-Gomez:2010jbn}. However,  for small-scales structures, this model suffers from the core-cusp problem, the too-big-to-fail problem and the diversity problem \cite{Oman:2015xda,Moore:1999gc,Tollerud:2014zha,Navarro:1996gj}.  

These small-scale anomalies may be solved via assuming colliding DM in different 
small-scale objects, such as galaxies and dwarf spheroidals, and this scenario is called the self-interacting DM (SIDM), which requires the DM self-scattering cross section per unit mass $\sigma/m_{{\rm DM}}$ to be about $(1-10)~ \rm{cm^2}/g$ \cite{Tulin:2017ara,Spergel:1999mh,Colquhoun:2020adl,Chu:2018faw,Chu:2018fzy,Tsai:2020vpi,Tulin:2012wi,Chu:2019awd,Buckley:2009in,Wang:2014kja,Kim:2022cpu,Kim:2021bmx,Wang:2022lxn,Garcia-Cely:2017qpx,Bringmann:2016din,Kahlhoefer:2017umn,Bernal:2015ova,Chu:2016pew}.  Generally, the feature of SIDM scenario is that the DM self-scattering cross section is velocity-dependent, associated with a light dark mediator, and usually has a small turn-over as going from constant scattering for the low-velocity dwarf spheroidals to $\sigma_Tm^2_{\chi} \propto v^{-4} $ for the high-velocity clusters of galaxies. As studied in ~\cite{Tulin:2013teo},  the accurate DM self-scattering cross section can be calculated via solving the Schr\"odinger equation using the partial wave analysis and can be classified as the Born, the resonant, and the classical regimes.  
 The most of parameter space preferred for solving the small-scale anomalies is found to be in the strongly coupled resonant and classical regions.  
In the quantum resonant regime, the cross section may have a non-trivial velocity dependence. In the resonance case it has $\sigma \propto v^{-2}$ while in the anti-resonance case it has $\sigma \propto v^0$.   The similar resonance behavior can also occur for the self-interacting puffy DM, in which the radius effect of DM particle can be another source of velocity dependence for $\sigma/m$. For these zero energy bound states, the cross section is enhanced strongly. Moreover, other non-perturbative effects of the boosted DM cross section may arise from the bound-state, the resonance,  the co-annihilation, and the Sommerfeld enhancement, which are relevant for the calculation of relic density and indirect detection of DM ~\cite{Wang:2022avs,Hisano:2004ds,Feng:2010zp,Arkani-Hamed:2008hhe,vonHarling:2014kha,Ellis:2018jyl}.  

As well known, the Sommerfeld enhancement was found by Arnold Sommerfeld in 1931~\cite{Sommerfeld:1931qaf}.  Before the low-velocity electron and positron annihilate into two photons, the Coulomb force between the two incoming particles may distort their wave-functions and thus the annihilation cross section may be affected. This non-relativistic quantum effect results in some correction to the scattering cross section between the low energy particles like the DM particles. From the perspective of quantum field theory, this effect can be described via summing over the ladder Feynman diagrams exchanging a massive dark mediator particle which can be denoted by an attractive Yukawa potential between a DM particle and an anti DM particle. In current particle physics, the Sommerfeld enhancement has been considered for the DM processes, such as the processes 
of freeze-in ~\cite{Zhong:2022mhw} or freeze-out DM ~\cite{Lee:1977ua}. The joint effect of resonant annihilation and Sommerfeld enhancement has been considered in the Higgs-portal scalar DM model and the simplified MSSM-inspired DM scenarios~\cite{Beneke:2022rjv}.  When the DM self-scattering process is near the zero-energy resonance, the Sommerfeld enhancement factor can be obtained by the Levinson's theorem and the effective range theory~\cite{Kamada:2023iol}. An analysis of the Sommerfeld enhancement to DM  annihilation in the presence of an excited state is provided in ~\cite{Slatyer:2009vg}.  Furthermore, this effect also provides a physics interpretation for the observed  cosmic  positron excess reported by PAMELA~\cite{PAMELA:2008gwm}, AMS-02~\cite{AMS:2019rhg}, and Fermi-LAT~\cite{Fermi-LAT:2011baq}. 

If the DM particle has a size, the self-interacting DM (SIDM) scenario may be readily realized and the composite SIDM may have dark atoms, nuclei and  bound states ~\cite{Laha:2013gva,Kondo:2022lgg,Cline:2021itd,Wang:2021tjf}. Recently, the study in ~\cite{Chu:2018faw} showed that the size effect of DM, in the presence of a light particle mediating the DM self-interaction, can solve the small-scale problems.  A further study on such puffy DM was performed via the partial wave analysis~\cite{Wang:2023xii}. In this work, we will first revisit the classification of self-scattering cross section for puffy DM via two new defined parameters. Then we examine the Sommerfeld enhancement for puffy DM processes. A key parameter for the self-interacting puffy DM, i.e., the ratio between $R_{\chi}$ (radius of a puffy DM particle) and  $1/m_{\phi}$ (force range), will be constrained with the consideration of Sommerfeld enhancement.

This paper is arranged as follows. In Sec. II, the particle physics dynamics for the self-interacting puffy DM is studied and the Sommerfeld enhancement for puffy DM is described.  In Sec. III, we constrain the self-interacting puffy DM with the consideration of Sommerfeld enhancement. Sec. IV gives our conclusions.

\section{Non-perturbative effect for puffy DM two-body system}
For the elastic scattering of two particles whose relativistic velocity is near the light speed, the perturbative effect is obviously dominant. As the velocity of the particles is dropping,  the non-perturbative effect will come into play. Especially, for the non-relativistic point-like DM self-scattering,  when the kinetic energy of the DM particles is very low,  this two-body system can form a quasi bound state in some range of potential which is wide and strong. In this case the scattering cross section will have a resonance enhancement. The same behavior can occur for the DM annihilation cross section,  which is called the Sommerfeld enhancement.  When the particle velocity becomes slower, the bound state of these two particles may be formed.  For the DM particle with a finite size, we also have such non-perturbative effect as studied in ~\cite{Wang:2023xii} where the cross section has a non-trivial velocity-dependence as for the point-like DM case~\cite{Tulin:2013teo}. In light of this, we may have Sommerfeld enhancement for puffy DM annihilation via a multi-exchange of the mediator $\rho$ or $\pi$ meson which forms an attractive Yukawa potential.

\begin{figure}[ht]
	\centering
	\includegraphics[width=8.cm]{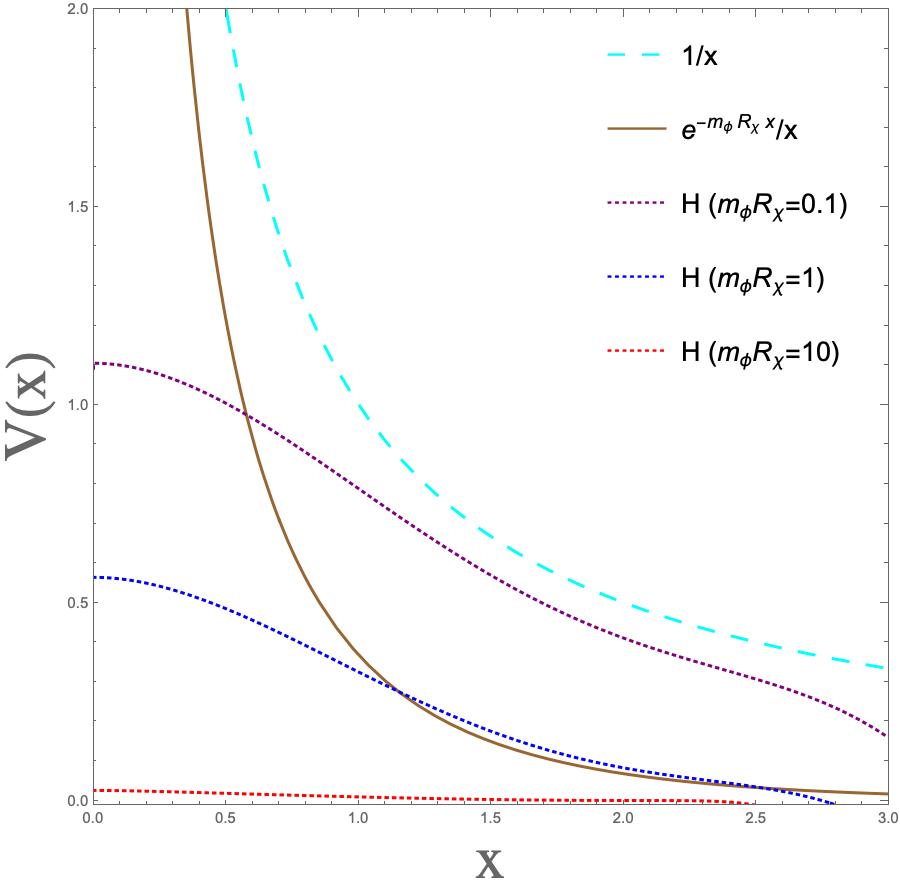}	
	\caption{A comparison of different potentials: the dashed curve is the Coulomb potential, the solid curve is the 
 point Yukawa potential ($m_\phi R_\chi=1$), while the three-dotted curves are the puffy potentials with $m_\phi R_\chi=0.1, 1, 10$ 
 for $r<2R_{\chi}$. }
	\label{fig1}
\end{figure}

First we recapitulate the dynamics of puffy DM self-scattering in the partial wave analysis, focusing on the classification of the cross section. According to two newly defined  parameters, the different regimes, namely the Born, the resonance and the classical, will be classified. Then the Sommerfeld enhancement for puffy DM $s$-wave annihilation will be explored. In such a puffy DM two-body system, the involved particles are the puffy DM particle (with mass $m_{\chi}$ and radius $R_{\chi}$) and the mediator particle (with mass $m_{\phi}$) which acts as a light force carrier and mediates an attractive interaction via the puffy potential 
\begin{align}\label{eq1}
V_{\rm puffy}(r) & = \begin{cases}
~- g(r,y) & r<2R_{\chi}\, , \\
\hspace{5cm}\  & \ \\[-6.mm]
~-\alpha\frac{e^{-m_{\phi}r}}{r} 
\times h\left(y\right)&  r>2R_{\chi}\,,
\end{cases}
\end{align}
where $y=R_{\chi}m_{\phi}$, $r$ is the relative distance between DM particles, and the dark fine structure constant $\alpha = g_{\chi}^2/4\pi$ with $g_{\chi}$ being the DM self-coupling. The specific form of $V_{\rm puffy}(r)$ is presented in Appendix~\ref{appa} and its derivation details can be found in Appendix A of ~\cite{Wang:2023xii}.  For $r<2R_{\chi}$, the puffy potential can be rewritten as a dimensionless form,  allowing for a comparison with the Yukawa and Coulomb potentials ~\cite{Wang:2023xii}:
\begin{align}\label{eq33}
V(r=R_{\chi}x) & = \begin{cases}
~\frac{\alpha}{R_{\chi}}
\frac{1}{x} & \rm Coulomb~ potential, \\
\hspace{5cm}\  & \ \\[-6.mm]
~ \frac{\alpha}{R_{\chi}}
\frac{e^{-yx}}{x} &  \rm Yukawa~ potential,\\
\hspace{5cm}\  & \ \\[-6.mm]
~  \frac{\alpha}{R_{\chi}}
H(x,y),& \rm Puffy~ potential, \\
\end{cases}
\end{align} 
where 
\be \label{Hp}
\begin{split}
	H(x,y)=&3\left\{y^4(-2+x)^3x(4+x)-6e^{-y(2+x)}\left[1+y+e^{2y}(-1+y)\right] \right.\\
	&\times \left[-2(1+y)+e^{yx}(2+y(2+(-2+y(-2+x))x))\right]\\
	&+6e^{-y}(1+y)(2(2+(y)^2(-2+x)x )\cosh(y)-4\cosh(y(-1+x)\\
	&\left.+4y(-1+x)\sinh(y)-\sinh(y(-1+x)) ))\right\}/(16y^6x).
\end{split}
\ee
As shown in Fig.\ref{fig1},  comparing with the point-like potential case, an important feature of this puffy potential is that for $r\rightarrow 0$, it no longer has a divergent pole due to the size effect.  So for the puffy DM, the solution of the Schr\"odinger equation and the associated non-perturbation effect such as the Sommerfeld effect may also be quite different from the point-like DM. 
%%%fig.1   
%%%%
In order to calculate these non-perturbative effects, we need to solve the Schr\"odinger equation for the puffy DM particle
\be\label{eq2}
-\frac{1}{2\mu}\nabla^2\psi_k(x)=\left(E-V_{\rm puffy}(r)\right)\psi_k(x),
\ee
where $\mu=m_{\chi}/2$ is the reduced mass, $E=k^2/2\mu$ with $k$ being the relative momentum.  When $r\rightarrow \infty$, the wave function has an asymptotic form 
\be\label{eq3}
\psi_k(x)\rightarrow e^{kz}+f(k,\theta)\frac{e^{ikr}}{r},
\ee
 where $f(k,\theta)$ is the scattering amplitude. The wave function using the partial-wave contributions can be decomposed as 
\be\label{eq4}
\psi_k(x)=\sum_l\frac{1}{k}e^{\frac{i}{2}l\pi+i\delta_l}\left(2l+1\right)R_{k,l}(r)
P_l(cos \theta),
\ee
where the phase shift $\delta_l$ of the $l$-th partial wave can be obtained by solving the Schr\"odinger equation for the radial wave function $\mathcal{R}_l(r)$  
\be\label{eq5}
\frac{1}{r^{2}}\frac{\partial}{\partial r}\left(r^{2}\frac{\partial\mathcal{R}_{k,l}}{\partial r}\right)+\left(k^2-2\mu V(r)-\frac{l(l+1)}{ r^{2}}\right)\mathcal{R}_{k,l}(r)=0.
\ee
The asymptotic solution for $\mathcal{R}_l(r)$ is given by   
\be \label{eq6}
\underset{r\rightarrow\infty}{\lim}\mathcal{R}_{l}(r)\propto\cos\delta_{l}j_{l}(kr)-\sin\delta_{l}n_{l}(kr),
\ee
where $j_l$ represents the spherical Bessel function and $n_l$ represents the spherical Neumann function.

\section{Self-scattering of puffy DM particles}\label{sec:SIDM}
Before discussing the dynamics of puffy DM, we introduce the condition of  Born approximation for puffy DM low-energy scattering,
\be\label{eq7}
m_{\chi}\left|\int_{0}^{\infty}rV_{\rm puffy}(r)dr\right|\ll1
\, ,
\ee
namely,
 \begin{eqnarray}
 m_\chi 
 \left|\int_{0}^{\infty}rV_{\rm puffy}(r)dr\right|
 &=& \frac{ \alpha m_{\chi}}{m_{\phi}}
 \frac{3(-15+y^2(15-10y+4y^3)+15(1+y^2)e^{-2y})}{10y^6}\nonumber\\
 &=& bf(y) \ll 1\,,
 \label{pufb}
 \end{eqnarray}
 where $b=\alpha m_{\chi}/m_{\phi}$  and a new function $f(y)$ is defined for the estimation of the validity. The new parameter $bf(y)$ plays an important role for the Born approximation, just as in the point DM case $b\ll1$ is the condition of Born approximation.  

Then, the self-scattering of two puffy DM particles generally has a transfer cross section 
\be\label{eq8}
\sigma_T=\int d\Omega(1-\cos\theta)d\sigma/d\Omega .
\ee 
Using the partial-wave approach, this transfer cross-section can be written as 
\be\label{eq9}
\frac{\sigma_{T}k^{2}}{4\pi}=\sum_{l=0}^{\infty}(l+1)\sin^{2}(\delta_{l+1}-\delta_{l}).
\ee
 In order to calculate the phase shift $\delta_l$,  the following  dimensionless parameters are defined:
 \be \label{eq10}
 \chi_l=rR_l, \quad  x=\alpha m_{\chi}r, \quad  a=\frac{v}{2\alpha},\quad  b=\frac{\alpha m_{\chi}}{m_{\phi}}\,. 
 \ee
In terms of these variables, the Schrödinger equation can be expressed as
\be\label{eq11}
\left(\frac{d^{2}}{dx^{2}}+a^{2}-\frac{l(l+1)}{x^{2}}-\frac{1}{m_{\chi}\alpha^{2}}V(r)\right)\chi_{l}=0\, .
\ee
Here the initial condition is set as $\chi_l(x_i)=1$ and $\chi'_l(x_i)=(l+1)/x_i$ 
and the point $x_i$ is near the origin. Then the  Schrödinger equation is solved within the range $x_i \leq x\leq x_m$, with $x_m$ being the maximum value of $x$ used in the numerical analysis. With the condition of asymptotic solution Eq.~(\ref{eq7}) and $x=x_m$, we have 
\be\label{eq12}
\chi_l\propto x e^{i\delta_{l}}(\cos\delta_lj_l(ax)-\sin\delta_ln_l(ax)).
\ee
Then the phase shift can be obtained  by
\be\label{eq13}
\tan\delta_l=\frac{ax_mj'_l(ax_m)-\beta_lj_l(ax_m)}{ax_mn'_l(ax_m)-\beta_ln_l(ax_m)},\quad \beta_l=\frac{x_m\chi'_l(x_m)}{\chi_l(x_m)}-1
\ee
The specific details of solving the Schr\"odinger equation can be found in Appendix B of Ref.~\cite{Wang:2023xii}.
%%%%fig.2
\begin{figure}[ht]
	\centering
	\includegraphics[width=7cm]{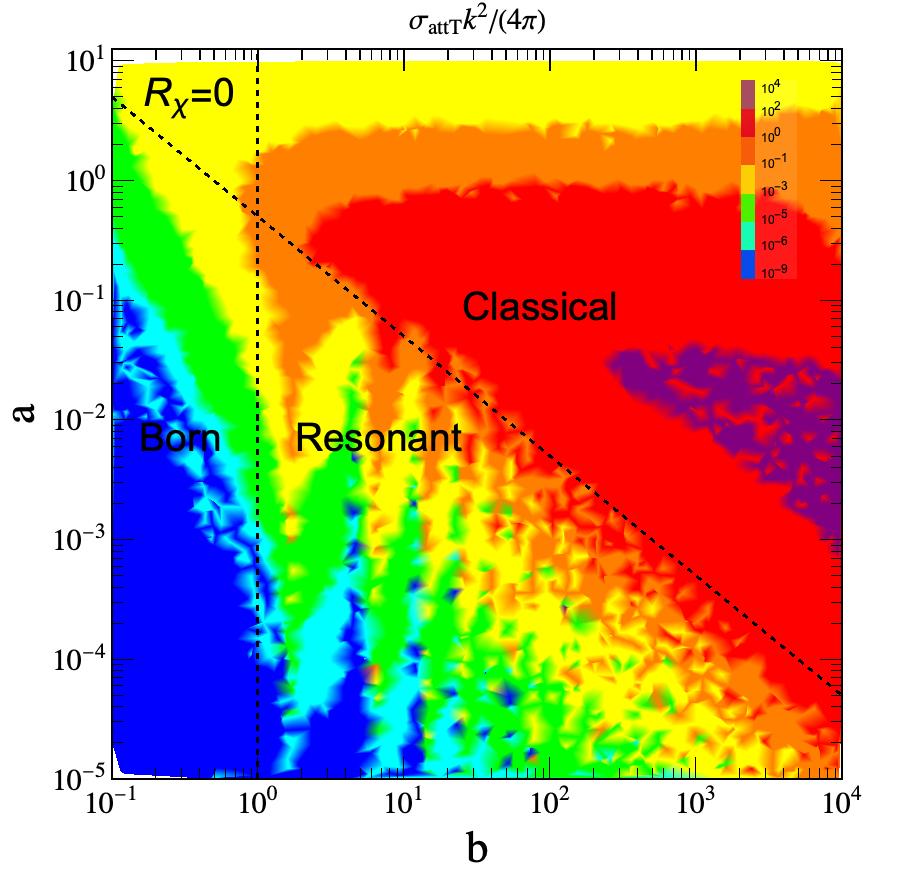}	
	\includegraphics[width=7.5cm]{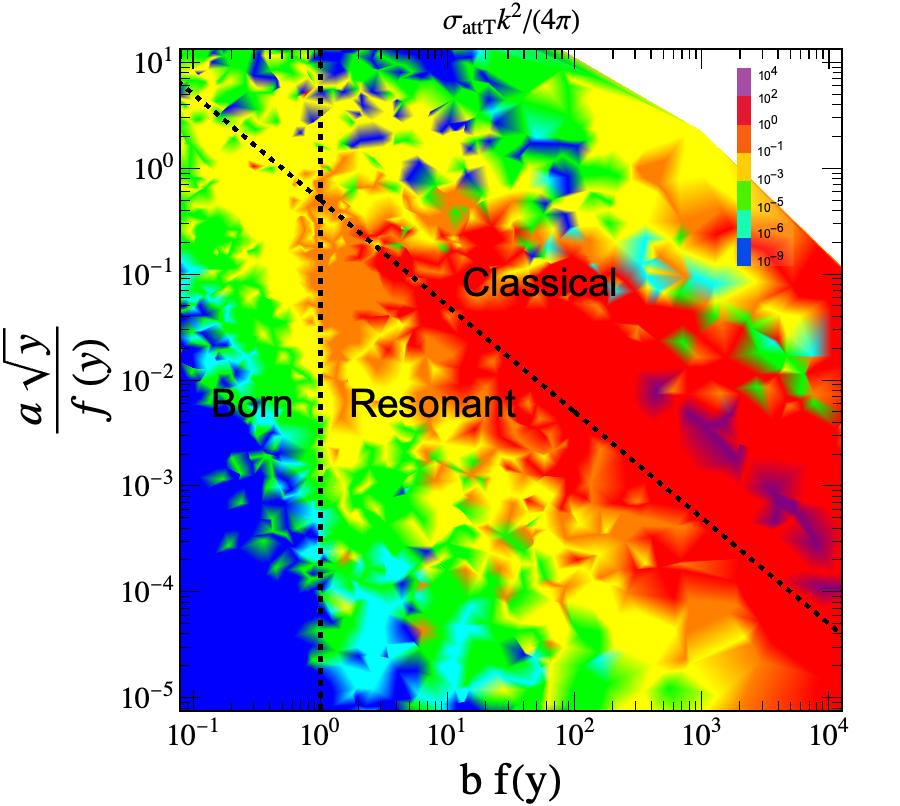}	
	\caption{The parameter space $(a,b)$ for point-like DM (left panel) and $(bf(y),a\sqrt{y}/f(y))$ for puffy DM (right panel), showing the value of $\sigma_Tk^2/(4\pi)$  with attractive force. The Born, classical and resonant parameter regimes are classified.}
	\label{fig2}
\end{figure}
%%% 
When $R_{\chi}=0$, the parameter space ($a,b$) with Yukawa potential $V(r)=-\alpha e^{-m_{\phi}r}/r$ is scanned. As shown in the left panel of Fig.\ref{fig2}, along with the  different value of  $\sigma_Tk^2/(4\pi)$,
the cross section of point-like DM  is divided into Born ($b<1$), resonant ($b>1$ and $m_{\chi}v/m_{\phi}<1$ ) and classical ($m_{\chi}v/m_{\phi}>1$) regimes. In the puffy DM case, we introduce two new parameters: $bf(y)$ and $\sqrt{y}a/f(y)$.  
Scanning over the parameter sapce ($m_{\chi},R_{\chi},\alpha,v,m_{\phi}$), we obtain the right panel of Fig.\ref{fig2} which shows the parameter space $(bf(y),\sqrt{y}a/f(y))$  and the corresponding value of $\sigma_Tk^2/(4\pi)$. So the cross section for puffy DM scattering is also shown as the Born, resonance and classical regimes:
\begin{itemize}
    \item The Born regime agrees with the condition of Born approximation $bf(y)<1$.  \item For the classical regime we have   
\be
\sqrt{m_{\chi}v/m_{\phi} \times m_{\chi}v R_{\chi}} ~> 1 .
\ee
\item For the resonance regime we have  $bf(y)>1$ and $\sqrt{m_{\chi}v/m_{\phi} \times m_{\chi}v R_{\chi}}<1$.  
\end{itemize}
Such a classification may make some physical sense and  have a profound influence for  the dynamics of puffy DM scattering. For example, for the classical regime,  the radius effect should be included besides the quantity $m_{\chi}v/m_{\phi}$  and can be expressed as  $m_{\chi}vR_{\chi}$ via replacing the force range $1/m_{\phi}$ as the radius $R_{\chi}$ of DM.  Thus, we can classify new regimes of puffy DM cross section in terms of the new dimensionless parameter $(m_{\chi}v) \times \sqrt{y}$ with the key parameter $y$ being the ratio between $R_{\chi}$ and the force range $1/m_{\phi}$. 

\section{Sommerfeld enhancement for puffy DM annihilation}\label{sec:scattering}
Now we study another non-perturbative effect for the puffy DM two-body system and derive the Sommerfeld enhancement factor for pufy DM annihilation. Supposing a  particle to move near some origin and we have a delta-form interaction to annihilate this particle. The rate of this process is proportional to the square of wave function $|\psi(0)|^2$. When the velocity of the incoming particle is low and the attractive central potential sizably distorts the wave function, then the annihilation cross section is boosted. Here we only consider the $s$-wave for puffy DM annihilation. 

The Sommerfeld enhancement factor can be written as 
\be\label{someq}
S=\frac{|\psi(0)|^2}{|\psi^0(0)|^2}=|\psi(0)|^2,
\ee
where $\psi^0$ is the $V=0$ wave function. This boost factor can be obtained by solving the Schr\"odinger equation Eq.~(\ref{eq4}). For accurate calculation, we define the variables
\be \label{eq16}
\chi_k=rR_{k,0}, \quad  x=m_\phi r, \quad  a=\frac{v}{2\alpha},\quad  b=\frac{\alpha m_{\chi}}{m_{\phi}}\, 
\ee
and then Eq.~(\ref{eq4}) can be expressed as
\be\label{eq17}
\frac{d\chi_k}{dx^2}=\left(\frac{m_{\chi}}{m_{\phi}^2}V(x)-(2ab)^2 \right)\chi_k.
\ee
The initial condition $\chi(0)=0$ and $\chi'(0)=1$ are set. Note that due to the linear Sch\"odinger equation, the wave solution of this initial condition is given as the $\tilde{\chi}$. Moreover,  the  different initial condition near the origin is set to obtain the value of phase shift $\delta_l$ of self-scattering cross section in  section \ref{sec:SIDM} and in this same condition, the  Sommerfeld enhancement factor also can be gotten as the Ref. \cite{Kamada:2023iol}.  The solution of original problem is given by assuming  $\chi=\tilde{\chi}/A$, with $\chi$ satisfying the primary boundary condition 
\be \label{eq18}
\chi(x\rightarrow \infty)\rightarrow \sin (2abx+\delta)
\ee
and $A$ being the asymptotic amplitude of $\bar{\chi}$. So, the Sommerfeld enhancement factor is 
\be 
S_k=\left| \frac{1}{k} \frac{d\chi_k}{dr}(0) \right|^2=\left| \frac{1}{2abA} \right|^2 .
\ee 
To calculate the amplitude $A$, we use the asymptotic wave form Eq. (\ref{eq18}) and obtain
\begin{eqnarray}
&&\tilde{\chi}=A\sin (2abx+\delta),\\ 
&& \frac{d\tilde{\chi}}{dx}=2abA\sin (2abx+\delta).
\end{eqnarray} 
Therefore, the amplitude A can be expressed as 
\be
A=\sqrt{\tilde{\chi}^2+\left( \frac{1}{2ab} \frac{d\tilde{\chi}}{dx} \right)^2}.
\ee 
%%%%fig.3 
\begin{figure}[ht]
	\centering
	\includegraphics[width=9cm]{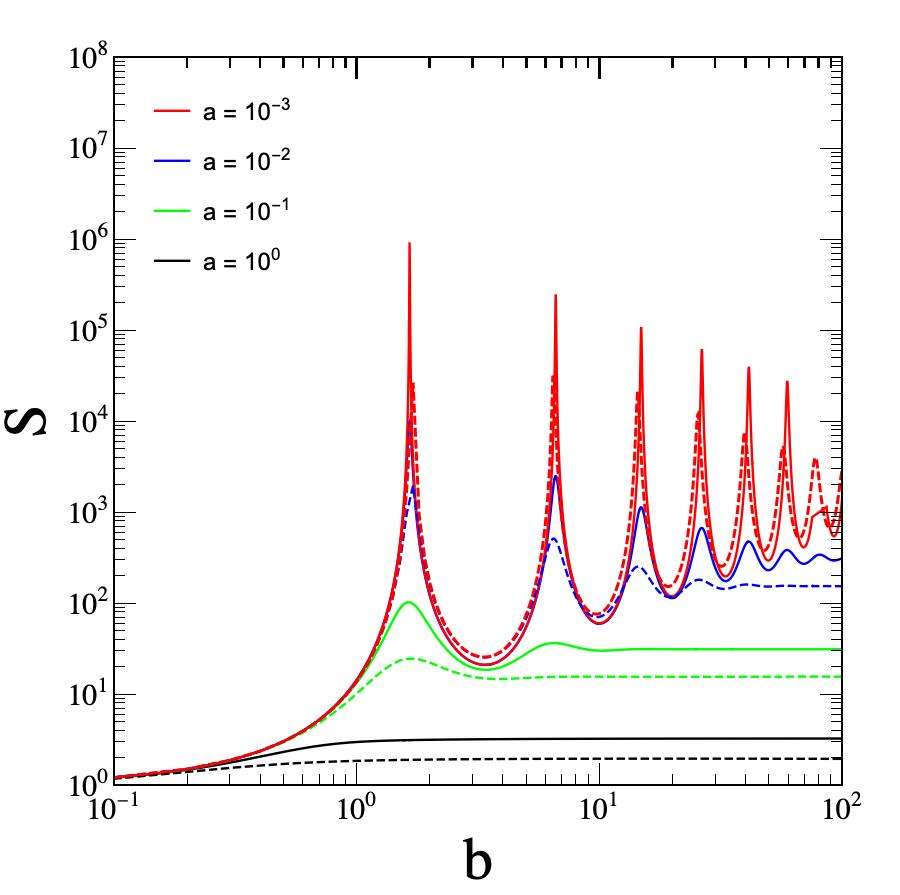}
	\caption{The Sommerfeld enhancement factor $S$ as a function of $b$ for the Yukawa potential. The dashed and solid curves denote the numerical and analytic results, respectively.}
	\label{fig3}
\end{figure}
To obtain the wave function $\tilde{\chi}$, the asymptotic value $x$ is solved numerically when the potential term is much less than the kinetic energy.  Then, we  firstly take the Yukawa potential for point-like DM into the Sch\"odinger equation Eq. (\ref{eq17}). The Sommerfeld enhancement factor as a function of $b$ is shown as the dashed curves in Fig. \ref{fig3}. The analytic expression of the Sommerfeld enhancement factor can be found by approximating the Yukawa potential as the Hulth\'{e}n potential:
\be
S=\frac{\pi}{10^a}\frac{\sinh \left(\frac{2\pi 10^a}{\pi^2 \frac{1}{6\times 10^b}}\right)}{\cosh \left( \frac{2\pi 10^a}{\pi^2 \frac{1}{6\times 10^b}}\right)-\cos \left( 2\pi \sqrt{\frac{1}{\pi^2 \frac{1}{6\times 10^b} }-\frac{\left(10^a\right)^2}{\left(\pi^2\frac{1}{6\times 10^b}\right)^2}}\right)}
\ee 
%%%fig.4 
\begin{figure}[ht]
	\centering
	\includegraphics[width=5.cm]{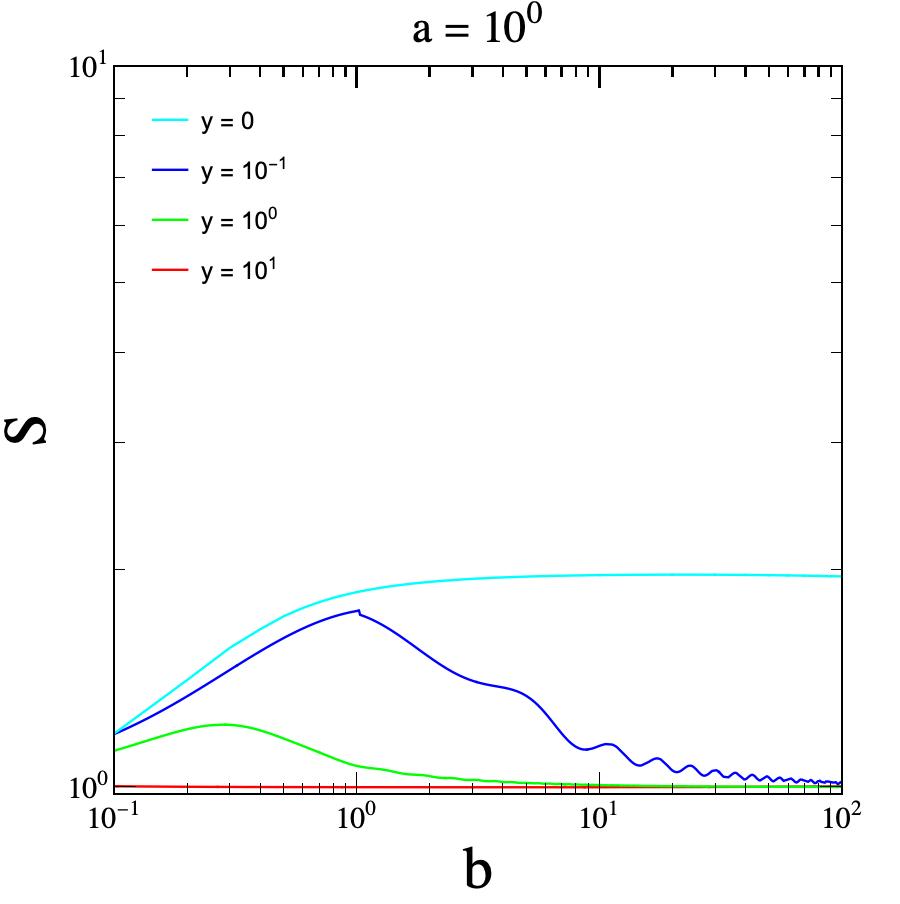}\hspace{-2mm}
	\includegraphics[width=4.9cm]{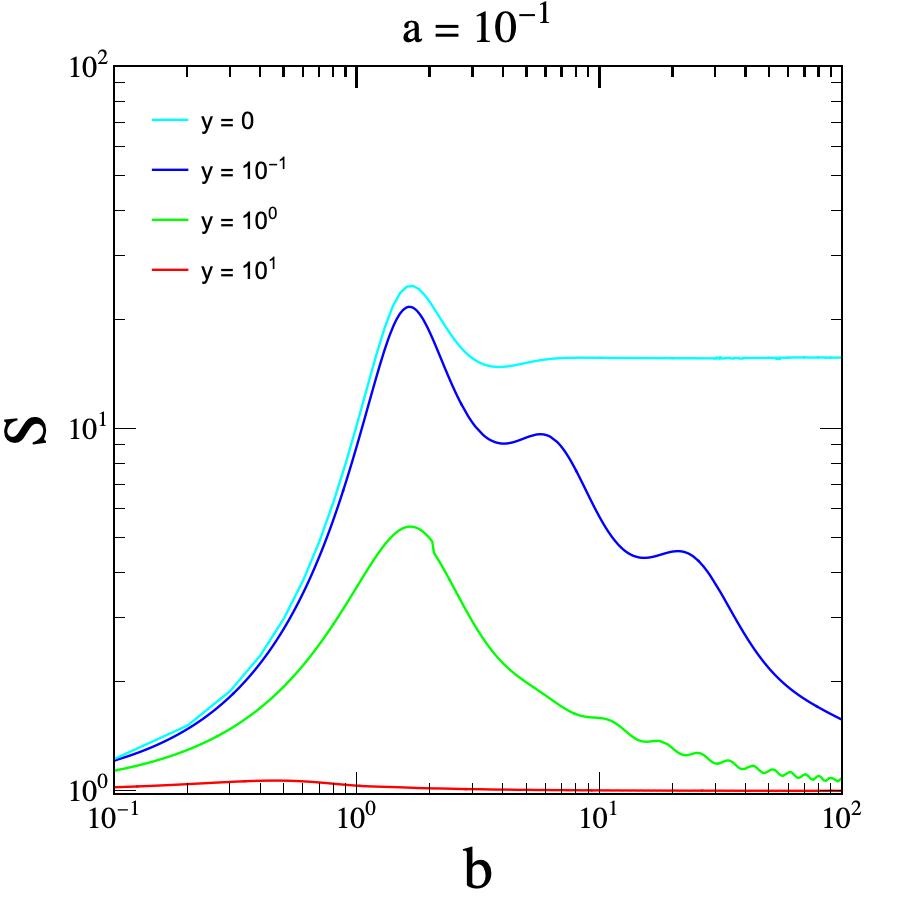}\hspace{-2mm}
	\includegraphics[width=4.9cm]{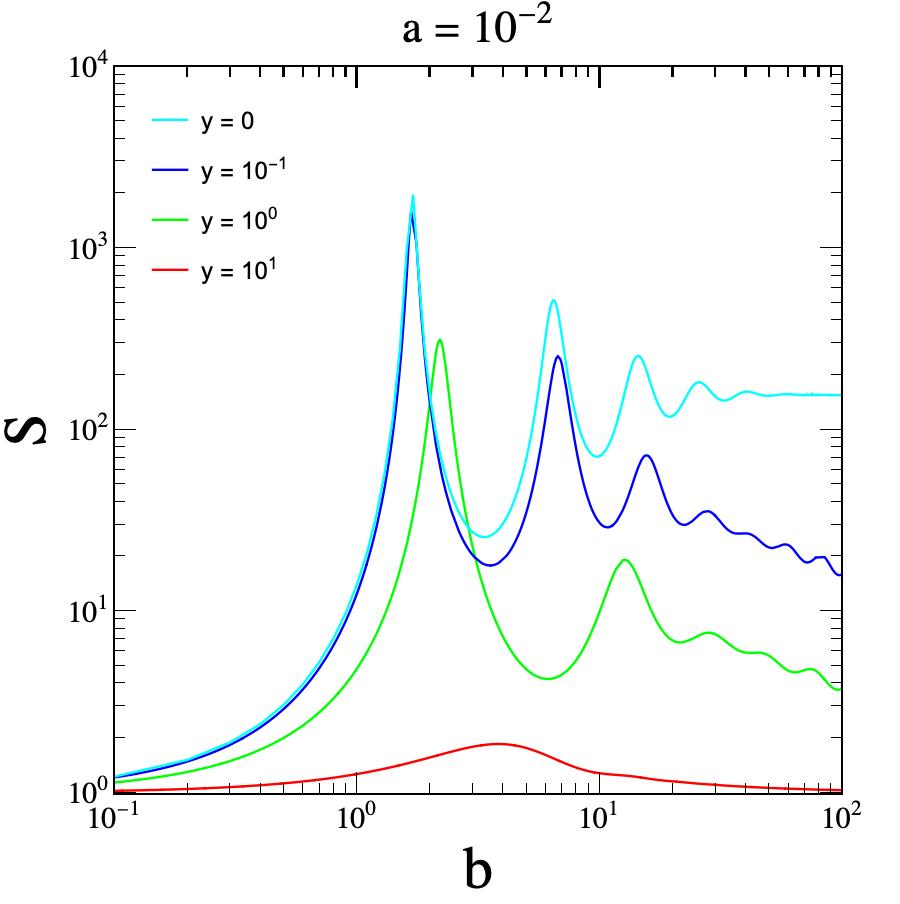}\\	
	\caption{The Sommerfeld enhancement factor $S$ of puffy potential as a function of $b$ for different values of $a$ and $y$.}
	\label{fig4}
\end{figure}
%%%%% 
Its result is shown as the solid curves in Fig.\ref{fig3}. With the decreasing velocity, the resonance becomes stronger. In the resonance region, the difference can be large between the analytic and numerical results. 

Next, we study the Sommerfeld enhancement for puffy potential case. Fig.\ref{fig4} shows that the Sommerfeld enhancement can happen for the puffy DM annihilation. And the dwarf galxies ($a=10^{-2}$) have a larger Sommerfeld enhancement of self-annihilating puffy DM cross section than the Milky way ($a=10^{-1}$) or clusters ($a=10^{0}$).  
When the parameter $a$ (proportional to the DM velocity) is decreasing, the cross section enhancement gets larger. With the increasing $b$, the Sommerfeld enhancement factor approaches to nearly unit and the resonance quantum effect tends to vanish.  The reason is that the initial values of the  puffy potential are almost constant but decrease continually and finally drops steeply as in Fig.\ref{fig1}. So for a large $b$, the potential will tend to zero due to the radius effect and the Sommerfeld enhancement will tend to vanish. Beyond that, when $a$ is fixed, the increasing value of $y$ can also lead to a weaker Sommerfeld enhancement and the reason is same as for a large $b$.  Therefore, the Sommerfeld enhancement can occur in the puffy DM annihilation. However, due to the vanishing divergent pole (compared with the point-like DM potential),  the Sommerfeld enhancement can become quite weak and the parameter space of the resonant state becomes narrower 
(compared with the point-like DM case). 

Next, we consider the thermal freeze-out with Sommerfeld enhancement for puffy DM \cite{vandenAarssen:2012ag,Liu:2023kat,Hisano:2006nn,Feng:2010zp,Arcadi:2017kky}.  The cosmological evolution of the abundance of puffy DM is described by the  Boltzmann equation:
\beq \label{Bolzeq}
\frac{dn_{\chi}}{dt}+3Hn_{\chi}=-\langle\sigma v\rangle(n_{\chi}^{2}-(n_{\chi}^{\rm eq})^{2}),
\eeq 
where the Hubble parameter $H=\dot{a}/a$ with $a$ being the scale factor,  $\langle\sigma v\rangle$ is the thermal average of annihilation cross section  and  $n_{\chi}$ is the number density of the puffy DM particles. In the non-relativistic limit,  the equilibrium number density is 
\beq\label{end}
n^{\rm eq}_{\chi}=g_{\chi}\left(\frac{ m_{\chi}^2T}{2\pi}\right)K_2\left(\frac{m_{\chi}}{T}\right),
\eeq 
where $g_{\chi}$ is the number of degree-of-freedom of the DM and   $K_2$ is the modified Bessel function of second order. For Eq. (\ref{Bolzeq}) we usually define the dimensionless variables:  
$Y_{\chi}(x)=n_{\chi}/s$, $Y_{{\chi},\rm eq}(x)=n_{{\chi}}^{\rm eq}/s$, $x=m_{\chi}/T$ with $s$ being the entropy density. When the annihilation rate is equal to the expansion rate of the universe ($\Gamma \sim H$), the relic density of DM can be obtained by the thermal freeze-out approach. The  freeze-out temperature can be determined by solving the equation:
\be 
\sqrt{\frac{\pi}{45}}M_{pl}\frac{g_{*}^{1/2}m_{\chi}}{x^{2}}\langle\sigma v\rangle Y_{\chi, \rm eq}\delta(\delta+2)=-\frac{d\log Y_{\chi,\rm eq}}{dx},
\ee 
where $\delta =(Y_{\chi}-Y_{\chi,\rm eq})/Y_{\chi,\rm eq}$ and $g_*$ is the effective relativistic degrees of freedom for energy density. In the present universe, an approximate solution is given by  
\be 
Y(T_{0})=Y_{0}=\sqrt{\frac{\pi}{45}}M_{pl}\left[\int_{T_{0}}^{T_{f}}g_{*}^{\frac{1}{2}}\langle\sigma v\rangle dT\right]^{-1}.
\ee 
The DM relic density can be described by the parameter $\Omega_{\chi}h^2$ which is the ratio between the DM energy density $\rho_{\chi}$  and the critical energy density $\rho_{\rm cr}$
\be 
\Omega_{\chi}h^{2}=\frac{\rho_{\chi}}{\rho_{c}}=m_{\chi}s_{0}Y_{0}\frac{h^{2}}{\rho_{cr}},
\ee 
with $\rho_{\chi}=m_{\chi}n_{\chi}=m_{\chi}s_{0}Y_{0}$ and $s_0$ being the entropy density at present time. Finally, the thermally averaged cross section is \cite{Gondolo:1990dk}
\be
\langle\sigma v\rangle=\frac{K_{2}(\frac{m_{\chi}}{T})^{-2}}{8m_{\chi}^{4}T}\int_{4m_{\chi}^{2}}^{\infty}ds\times\sigma(s)\sqrt{s}(s-4m_{\chi}^{2})K_{1}(\frac{\sqrt{s}}{T}).
\ee

\begin{figure}[ht]
	\centering
	\includegraphics[width=9cm]{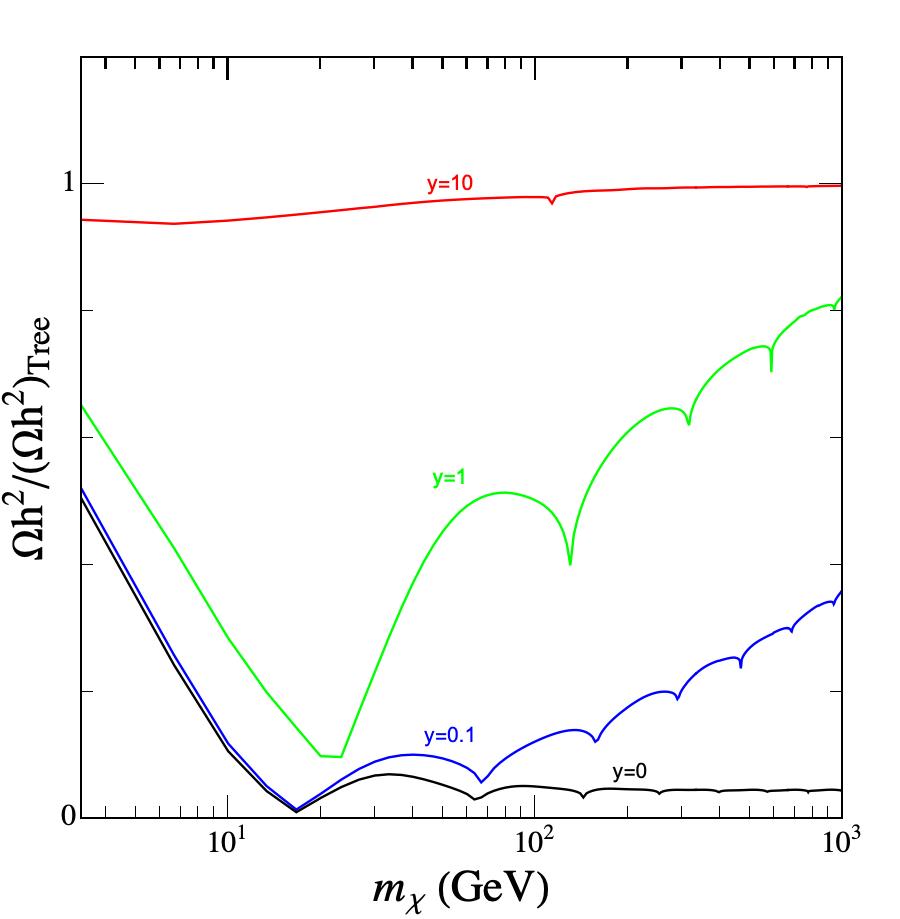}
	\caption{ The Sommerfeld effect for the relic density of pufffy DM ($y=0.1, 1, 10$) and point-like DM ($y=0$). Here $\alpha=0.1$ and $m_{\phi}=1~ \rm GeV$ are taken.}
	\label{fig5}
\end{figure}

In our study, for simplicity, the annihilation channel $\chi \bar{\chi}\rightarrow \phi\phi$ is considered and  the tree-level cross section is given as $\langle\sigma v\rangle_0 \simeq \pi \alpha^2/m^2_{\chi}$. For the $s$-wave, the annihilation cross section with  Sommerfeld enhancement is 
\be
\langle\sigma v\rangle= \bar{S}(v) \times \langle\sigma v\rangle_0,
\ee 
where 
\be
\bar{S}=\frac{K_{2}(\frac{m_{\chi}}{T})^{-2}}{8m_{\chi}^{4}T}\int_{4m_{\chi}^{2}}^{\infty}ds\times S(s)\sqrt{s}(s-4m_{\chi}^{2})K_{1}(\frac{\sqrt{s}}{T}). 
\ee
In order to study the Sommerfeld effect in puffy DM relic density, we take different values $y=10, 1,0.1$  and use the public code DRAKE to calculate the relic density \cite{Binder:2021bmg}.  As shown in Fig.\ref{fig5}, the four curves indicate that the Sommerfeld effect for puffy DM can affect the relic density due to the size effect. For the point-like DM case (the black curve in  Fig.\ref{fig5}), the Sommerfeld effect is stronger than the puffy DM case.  When the value of $y$ increases,  the Sommerfled effect becomes weaker. For $y=10$, the Sommerfled effect almost vanishes, i.e., $\Omega h^2/(\Omega h^2)_{\rm tree} \sim 1$.
%%%%fig.5 

\section{Puffy SIDM scenario and Sommerfeld enhancement}\label{sec:cons}
\begin{figure}[ht]
	\centering
	\includegraphics[width=4.9cm]{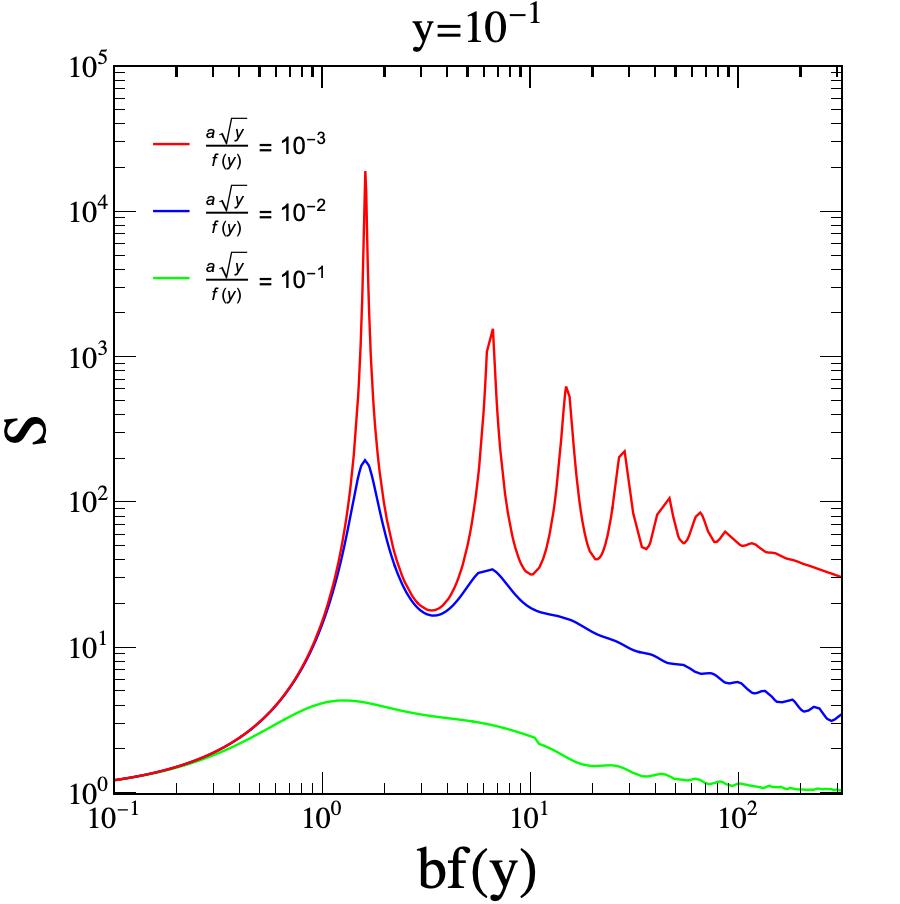}\hspace{-2mm}
	\includegraphics[width=4.9cm]{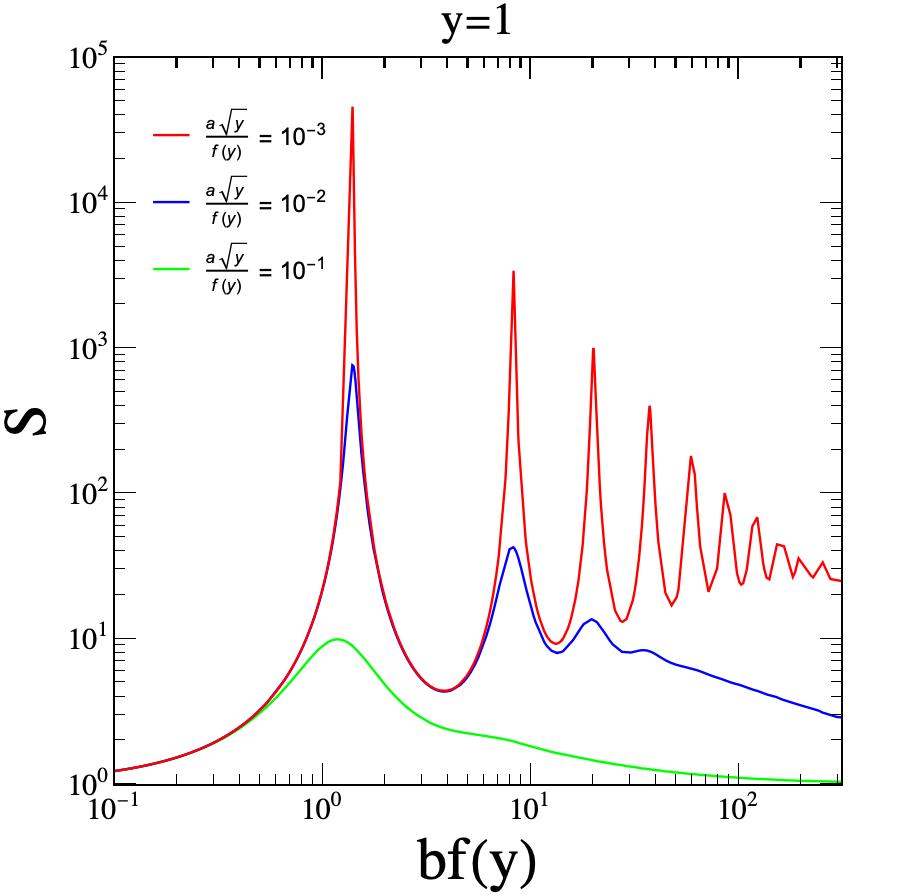}\hspace{-2mm}
	\includegraphics[width=4.9cm]{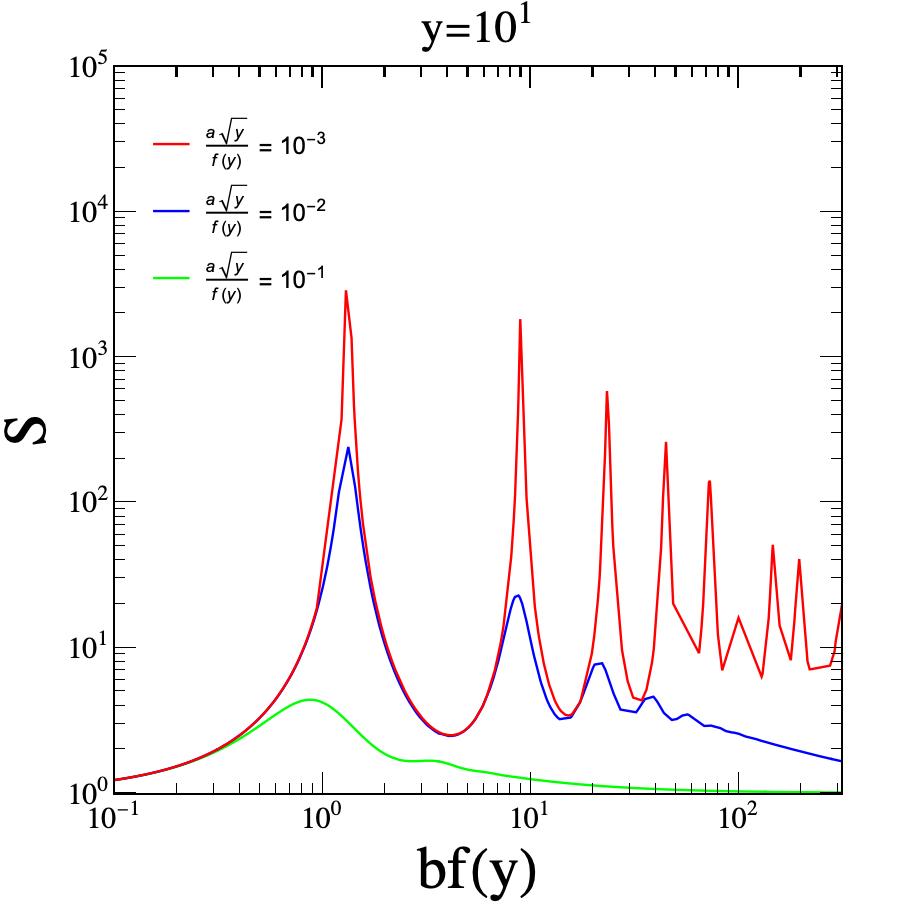}\\	
	 \vspace{5mm}
	\includegraphics[width=4.9cm]{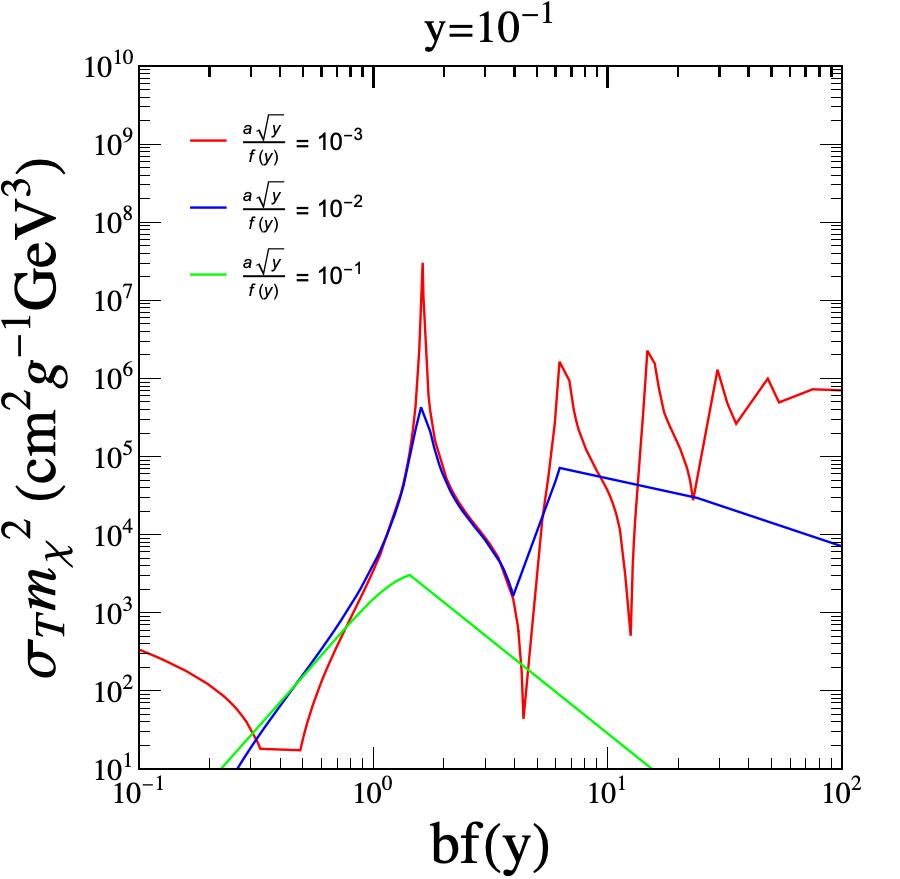}	\hspace{-2mm}
	\includegraphics[width=4.9cm]{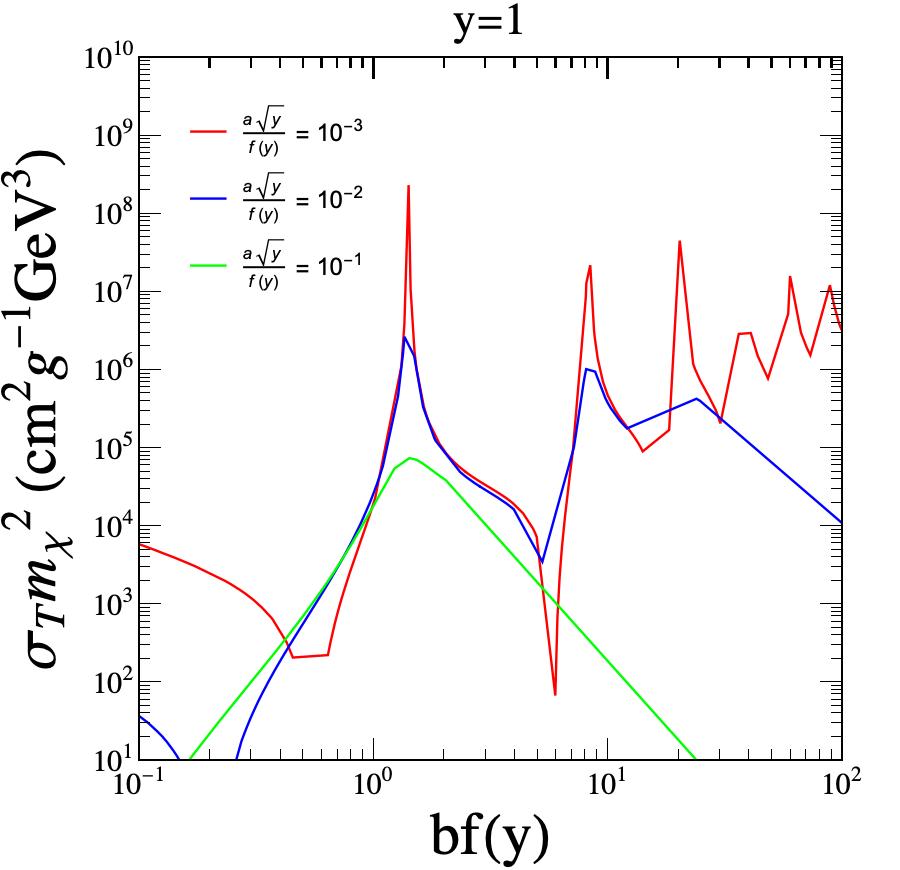}\hspace{-2mm}
	\includegraphics[width=4.9cm]{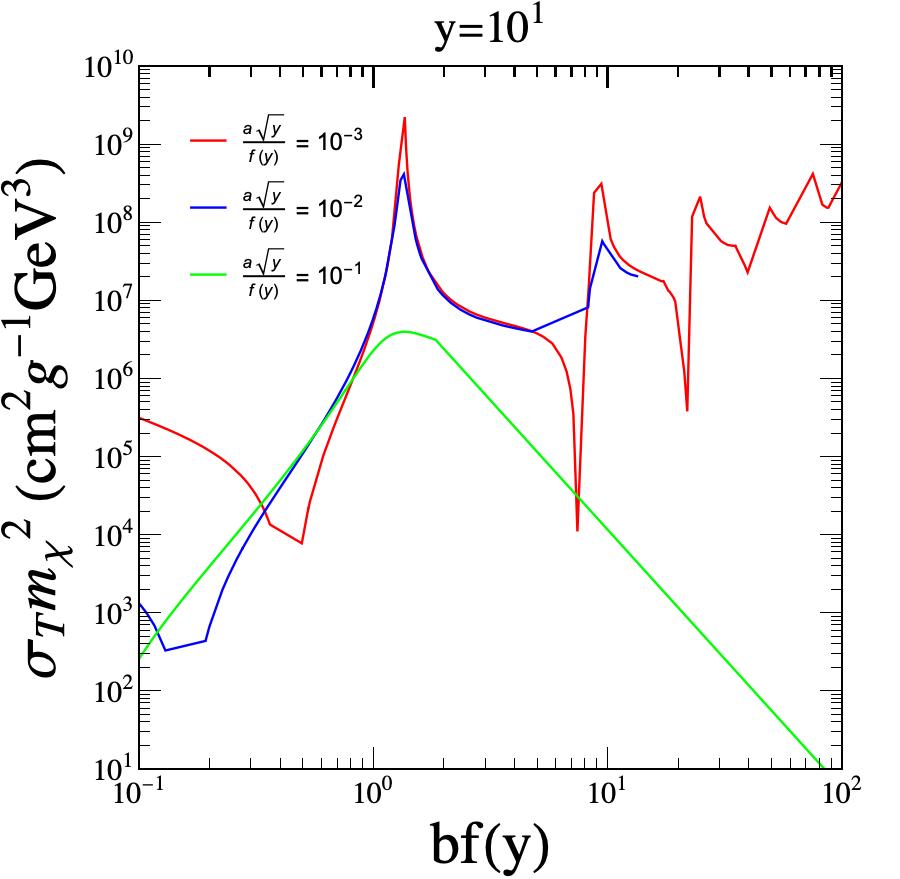}\\	
	\caption{Puffy DM self-interacting cross section $\sigma_Tm_{\chi}^2$ (lower panel) and the Sommerfeld enhancement factor $S$ (uppder panel) for the $s$-wave solution with different values of $y$ and $a\sqrt{y}/f(y)$.}
	\label{fig6}
\end{figure}

In the point-like DM case, both the velocity-dependent self-interacting cross section and the Sommerfeld enhancement for DM annihilation cross section are obtained from solving the same Schordinger equation, whose $s$-wave solution has the resonance behaviour of the zero-energy bound state. For the $s$-wave solution, by approximating the Yukawa potential as the Hulth\'{e}n potential, the resonance conditions   
$b/(\pi^2/6)=n^2 ~(n=1,2,3, \cdots)$ can be obtained.  
We expect that such resonance behaviours also occur for both the puffy SIDM cross section and the Sommerfeld enhancement in terms of the new parameters $bf(y)$ and $a\sqrt{y}/f(y)$ although the analytic resonance conditions are hard to derive.  Fig.\ref{fig6} shows that for different values of $y$ (taken as $10^{-1}$, $1$ and $10^1$) and $a\sqrt{y}/f(y)$ (taken as $10^{-1}$, $10^{-2}$ and $10^{-3}$), the  resonances or anti-resonances indeed happen at the same locations for both the self-interacting cross section $\sigma_Tm_{\chi}^2$ and the Sommerfeld enhancement factor $S$ for the s-wave situation. It further indicates that the same bound state formation is relevant for puffy DM scattering and annihilation.
%%%fig.6 
%%%%%%%%%%%%%

The puffy SIDM may generate different signals from the point-like SIDM in direct detection experiments due to the radius effect. The Sommerfeld effect should also be considered in the indirect detections. In the following, we consider the puffy SIDM scenario with the Sommerfeld enhancement. For the puffy DM, as shown in our previous work, the parameter $y=R_\chi m_\phi$ (the ratio between the radius of DM particle and the mediator force range) is a key parameter. Thus, we examine the parameter space ($m_{\chi}, y$). 
%%%fig.7 
\begin{figure}[ht]
	\centering
	\includegraphics[width=7.4cm]{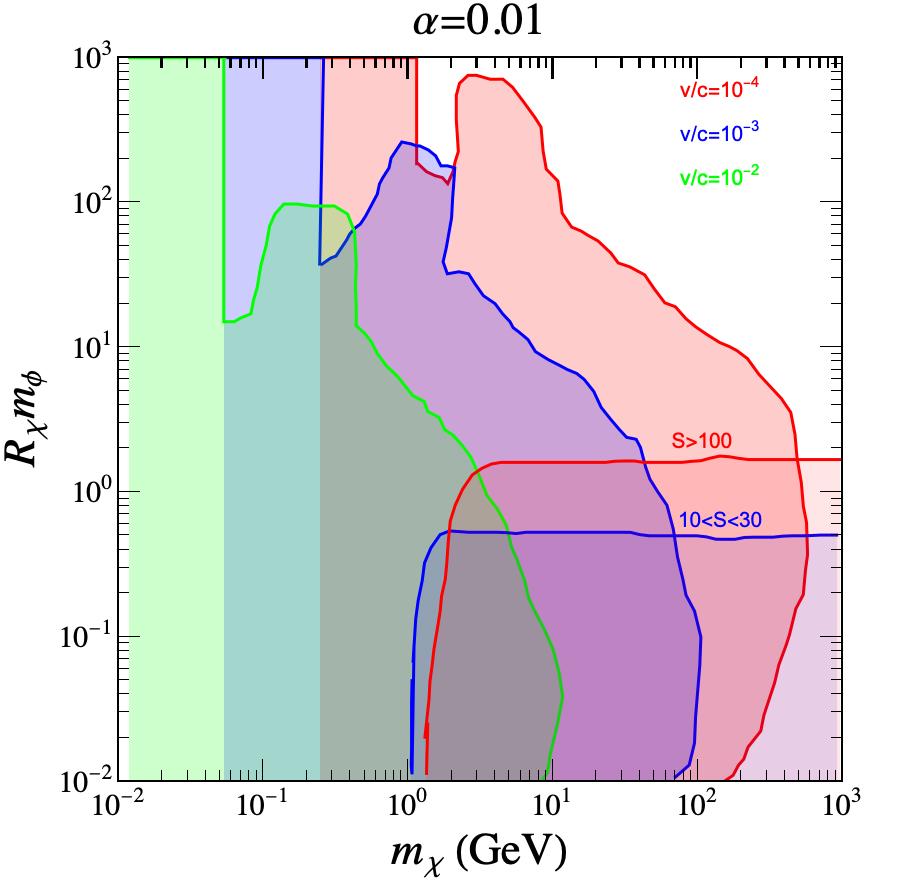}	
	\includegraphics[width=7.4cm]{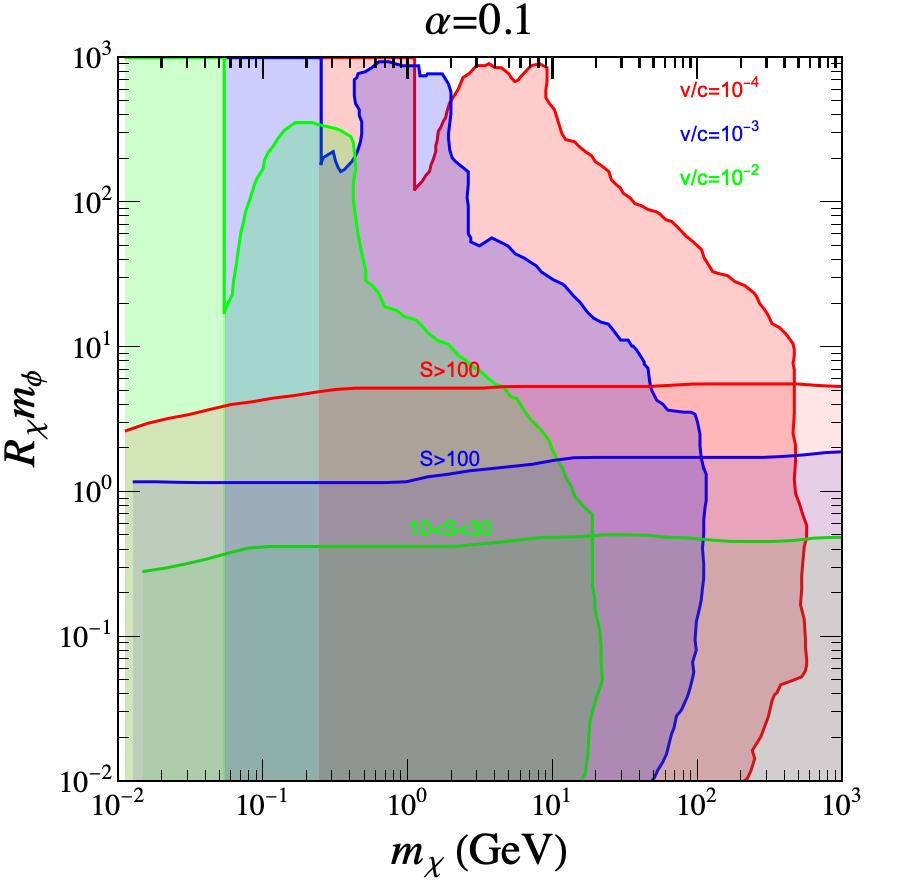}	
	\caption{The parameter space  ($m_{\chi}, R_\chi m_\phi$) for the puffy SIDM self-scattering cross section per unit mass  $0.1~{\rm cm^2/g}<\sigma_T/m_{\chi}<10~ \rm{cm^2/g}$ 
 to solve the small-scale problems. 
The red, blue and green regions correspond to the solutions for different small cosmological scales with a fixed velocity $v/c=10^{-4}$, $10^{-3}$ and $10^{-2}$, respectively (corresponding to the dwarf galaxies, Milky Way galaxy and cluster scales, respectively).
 The values of the Sommerfeld enhancement factor are also displayed (for example, the bottom-right corner region surrounded by the red curve labelled with $S>100$ in the left panel and the region below the red curve labelled with $S>100$ in the right panel have $S>100$ for $v/c=10^{-4}$). The $S$ factor for $v/c=10^{-2}$ in the left panel approaches to 1 and is not displayed.}  
	\label{fig7}
\end{figure}
%%%%

In our calculation the coupling constant is fixed at $0.01$ or $0.1$. In order to solve the small-scale problem, Fig.\ref{fig7} shows that the parameter space ($m_{\chi}, y$) for the puffy SIDM cross section per unite mass $0.1~{\rm cm^2/g}<\sigma_T/m_{\chi}<10~ \rm{cm^2/g}$ at different small cosmological scales with a fixed velocity $v/c=10^{-4}$,  $10^{-3}$ and  $10^{-2}$ corresponding to the dwarf galaxies, Milky Way galaxy and cluster scales, respectively. Each rectangular region corresponds to the Born regime of self-interaction, followed (from left to right) by the resonant or classical regime. Comparing the two panels of Fig.\ref{fig7}, a bigger coupling  constant $\alpha$ can give a larger region with a large Sommerfeld enhancement factor. When $a=v/2\alpha \sim 1$, the Sommerfeld effect vanishes as shown in the left panel of Fig.\ref{fig4}, so that for $v\sim 0.01$ in the left panel the Sommerfeld enhancement factor $S\sim 1$.  Note that a large $\alpha$ may lead to a large DM-nucleon scattering cross section which may be in conflict with the direct detection limits.  

\section{Conclusion}\label{sec:con}
For puffy DM, the size effect of the DM particle leads to the disappearance of the Yukawa potential pole and correspondingly the influence of quantum non-perturbative effects will have a great change. In this work we focused on the puffy DM and  discussed the reclassification of scattering cross sections and the Sommerfeld enhancement for puffy DM. We obtained the following observations: (i) The Sommerfeld enhancement factor approaches to 1 (no enhancement) for a large size with $y=10$; (ii) The value of the parameter $b$ (corresponding to the maximum resonant value) increases with the value of $y$; (iii) The resonance behaviors of the Sommerfeld enhancement for the puffy DM annihilation cross section and for the self-interacting cross sections have the same locations in terms of the new parameters $a\sqrt{y}/f(y)$ and $bf(y)$ to reclassify the self-interacting cross sections. Finally, for the puffy SIDM scenario to solve the small-scale problems, the values of the Sommerfeld enhancement
factor are displayed in the allowed parameter regions.

\section*{Acknowledgements}
W.-L. Xu thanks Xingchen Xu  for useful discussions about the  zero-energy bound state. This work was supported by the Natural Science Foundation of China (NSFC) under grant numbers 12075300, 11821505, and 12275232, the Peng-Huan-Wu Theoretical Physics Innovation Center (12047503), and the CAS Center for Excellence in Particle Physics (CCEPP).

\appendix
\section{The puffy Yukawa potential}\label{appa}
The Yukawa potential among two spheres is given by 
\begin{align}\label{eqa}
V_{\rm puffy}(r) & = \begin{cases}
\pm g(y) & r<2R_{\chi} \\
\hspace{5cm}\  & \ \\[-6.mm]
\pm\alpha\frac{e^{-m_{\phi}r}}{r} \times h\left(y\right)&  r>2R_{\chi},
\end{cases}
\end{align}
where
\begin{eqnarray} \label{inbol}
	g(r,y)&=&-\alpha4\pi^{2}(\frac{3}{4\pi R_{\chi}^{3}})^{2}\frac{(1+y)e^{-y}}{m_{\phi^{3}}}\frac{1}{2m_{\phi}^{3}} \nonumber \\
	&& \times\left(\frac{2e^{-m_{\phi}(R_{\chi}+r)}(-1+e^{y})(e^{2y}(-1+y)+e^{y}(1+y))}{r}\right.\nonumber \\
	&& \Bigg.+m_{\phi}(-e^{y}(2+m_{\phi}(r-2R_{\chi}))+e^{-y}(2-m_{\phi}r+2y))\Bigg)\nonumber \\
	&& +\alpha4\pi^{2}\left(\frac{3}{4\pi R_{\chi}^{3}}\right)^{2}\frac{1}{m_{\phi}}\left(\frac{e^{-y}}{m_{\phi}^{2}}-\frac{e^{y}}{m_{\phi}^{2}}+\frac{R_{\chi}e^{-y}}{m_{\phi}}+\frac{R_{\chi}e^{y}}{m_{\phi}}\right) \nonumber \\
	&& \times\frac{e^{-m_{\phi}(R_{\chi}+r)}(2+2y-e^{y}(2+m_{\phi}r(-2+m_{\phi}(r-2R_{\chi}))+2y))}{2m_{\phi}^{3}r}\nonumber  \\
	&& +\alpha4\pi^{2}\left(\frac{3}{4\pi R_{\chi}^{3}}\right)^{2}\frac{1}{m_{\phi}^2}\left( \frac{1}{12}(r-2R_{\chi})(r+4R_{\chi})\right), \\
	h\left(y\right)&=&4\pi\left(\frac{3}{4\pi R_{\chi}^{3}}\right)^{2}\frac{\pi}{m_{\phi}^{2}}\left(\frac{e^{-y}}{m_{\phi}^{2}}-\frac{e^{y}}{m_{\phi}^{2}}+\frac{R_{\chi}e^{-y}}{m_{\phi}}+\frac{R_{\chi}e^{y}}{m_{\phi}}\right)^{2}
\end{eqnarray}

\bibliographystyle{apsrev}
\bibliography{note}

\end{document}